# The Role of Gyrating Ions in Reformation of a Quasi-parallel Supercritical Shock

Hadi Madanian[1] and Terry Z. Liu[2]

[1]Epex Scientific, Richmond, VA 23294, USA (hadi@epexsci.com)

[2]University of California Los Angeles, Los Angeles, CA 90095, USA

## Abstract

Collisionless shocks in space and astrophysical plasmas mediate energy exchange between charged particles and fields in two or more plasma flows. In this study we analyze the evolution of ion distributions around a reformation cycle of a quasi-parallel shock. We use multi-point in-situ observations in the foreshock region of the Earth's bow shock of a transient foreshock structure as it generates a shock. We find that backstreaming ions in the foreshock create a density and magnetic field depletion known as caviton which locally changes the shock geometry. Gyrating suprathermal ions that emerge within the caviton and reach the upstream edge of the core create a cross-field current imbalance that results in the nonlinear growth of a new shock layer. The new shock forms from the background foreshock fields over a distance of ~ 6 ion inertial lengths ($l_i$) and within 4.5 to 11.2 $l_i$ from the main bow shock. We find that plasma compression at the new thin shock layer is due to compactification of the cold upstream ion beam by high amplitude magnetic field-aligned electrostatic fields. At later stages, the plasma compression expands to form a new sheath.

## 1 Introduction

Formation and propagation of shocks in collisionless plasmas is largely influenced by the dynamics of charged particles across the shock. The magnetic field in the upstream region plays an important role in the charged particle dynamics. At quasi-parallel shocks, where the angle between the upstream magnetic field and the shock propagation direction is small (i.e., below ~45°), some incident ions can be reflected to stream back and travel far distances along the upstream magnetic fields in the foreshock region (Eastwood et al., 2005). Commonly observed distributions upstream of planetary bow shocks include the field-aligned beam (FAB) ions which are gyrotropic ion distributions propagating upstream at small angles to the magnetic field, gyrating ions that also propagate upstream at different pitch angles, gyrophase bunched ions that can be classified as non-gyrotropic FAB ions, intermediate kidney-shaped distributions, and isotropic diffuse ion distributions (Fuselier, 1995; Kajdič et al., 2017; Kempf et al., 2015;

Savoini & Lembège, 2015; Stasiewicz & Kłos, 2022; Trattner et al., 2023). Backstreaming ions introduce a new source of energy in the foreshock and depending on the particle distribution and resonance conditions in the ambient plasma, plasma instabilities and perturbations grow in the upstream region giving rise to foreshock waves (Gary, 1991). FAB ions are commonly observed with ultra-low frequency (ULF) waves in the foreshock region (Hoppe et al., 1981), but FAB ions are insufficient to drive high amplitude steepened magnetic perturbations (Omidi et al., 1994; Wilson et al., 2013). With increase in the flux of suprathermal ions closer to the bow shock, the growth rate of foreshock transient shocks and associated pressure variations increases (An et al., 2020; Tarvus et al., 2021). Within the ULF foreshock region, nonlinear steepening and growth of quasi-periodic ULF waves upstream of quasi-parallel shocks leads to short large amplitude magnetic structures (SLAMS) (Johlander et al., 2022; Schwartz et al., 1992). SLAMS propagate upstream but are convected by the solar wind towards the bow shock initiating reformation cycles. The quasi-periodicity of SLAMS and reformation cycles is driven by the ion cyclotron frequency of the backstreaming protons that cause the underlying instability.

A quasi-parallel shock also undergoes reformation due to nonuniform reflection across a rippled shock surface where cold reflected ions interact with foreshock waves to generate a new shock (Hao et al., 2017). Backstreaming ions also evolves as they propagate upstream and interact with foreshock waves and the foreshock can contain a mixture of various ion populations (Meziane et al., 2004; Winske & Leroy, 1984). Nonlinear interaction of ULF wave modes can result in foreshock cavities or ``cavitons'' that manifest as correlated decreases in magnetic field strength and plasma density within the ULF foreshock region (Blanco-Cano et al., 2009). Similar depleted structures can form in the foreshock by the increased thermal pressure due to suprathermal ions which modify the incident solar wind pressure and excavate cavities in the plasma and the magnetic fields of the foreshock (i.e., hot diamagnetic cavities) (Parks et al., 2006; Sibeck et al., 2002; Thomsen et al., 1986). The depleted plasma and field in this case are accumulated near the cavity rims forming the cavity boundaries and distinguishes these structures from magnetic holes carried by the solar wind which are typically pressure balanced structures (Madanian et al., 2022). A spontaneous hot flow anomaly (SHFA) is a foreshock transient structure that can grow out of a caviton as it further interacts with backstreaming foreshock ions (Omidi, Zhang, et al., 2013; Zhang et al., 2013). Depending on the spacecraft position across the bow shock, foreshock transient features can appear in different orders in timeseries data (Omidi, Sibeck, et al., 2013). Most SHFAs tend to propagate perpendicular to the background magnetic field (Zhu et al., 2021) and are characterized by a depleted core and a compressive shock layer on one or both sides of the core. For some foreshock transients associated with upstream discontinuities, a rather nonmagnetized motion of backstreaming suprathermal ions, with respect to the magnetized electrons, across the current sheet gives rise to Hall currents that shape up the foreshock transient structures (Liu et al., 2020; Vu et al., 2023). Once formed, foreshock transient shocks can develop their own shock microstructures through cross scale ion-electron kinetic processes (Liu et al., 2016; Turner et al., 2021; Wang et al., 2022).

Upstream perturbations and cyclic reformation at quasi-parallel shocks scatter and accelerate foreshock ions making them available for further energization processes and generation of energetic particles such as solar energetic particles or galactic cosmic rays. In addition, pressure variations associated with foreshock structures has important implications for space weather. It is important to identify processes responsible for shock generation and to determine the foreshock ion dynamics at different stages of formation. In this study, we analyze the generation of a supercritical shock layer in the foreshock region of the Earth's quasi-parallel bow shock which generates its own sheath and eventually replaces the bow shock. We use multi-point spacecraft observations to determine plasma conditions and ion distributions at various phases of formation and to characterize how foreshock ions and the new shock layer evolve during the reformation cycle. Data sources and observations are shown in Section 2, results are discussed in Section 3, and Conclusions are provided in Section 4.

## 2 Data Sources and Observations

We focus on a foreshock transient event upstream of Earth's bow shock on 10 February 2025 at near 04:48:00 UT observed by the Magnetospheric Multiscale (MMS) spacecraft constellation in a string-of-pearl configuration. This event is observed by three MMS spacecraft on an outbound/sunward trajectory. Figure 1a shows the geospatial position of the probes with respect to a modeled bow shock, while the formation of the three MMS spacecraft is shown in Figure 1b. Timeseries data in panels c – n in Figure 1 show MMS1, MMS2, and MMS3 observations of the event. MMS4 is several $R_E$ downstream in the magnetosheath and is not considered. MMS1 is positioned at ~ (12.0, -2.0, -6.7) $R_E$ moving ahead of MMS3 and MMS2. The magnetic field data in panel Figure 1c shows foreshock perturbations and ULF wave activities most notably at ~ 1 Hz. The bow shock angle is $\theta_{Bn}$ = 43°, using the shock normal vector $\hat{n}$=(0.95, -0.11, -0.29) obtained from the mixed coplanarity method (Schwartz, 1998), and using MMS probe 3 (MMS3) data across the main bow shock. This estimate is very close to the shock normal direction based on a conic section model fitted to the position of MMS1. The upstream plasma fast magnetosonic Mach number is $M_f$ = 6.2 and the ion plasma beta of $\beta_i$ = 0.2. MMS1 and

MMS3 are separated by 6.5 $l_i$, while the distance between MMS3 and MMS2 is 2 $l_i$, where $l_i$ is the upstream ion inertial length ~ 115 km. The three MMS spacecraft are also roughly aligned along the bow shock normal which allows for analyzing the foreshock structure at different stages of formation and different distances from the bow shock.

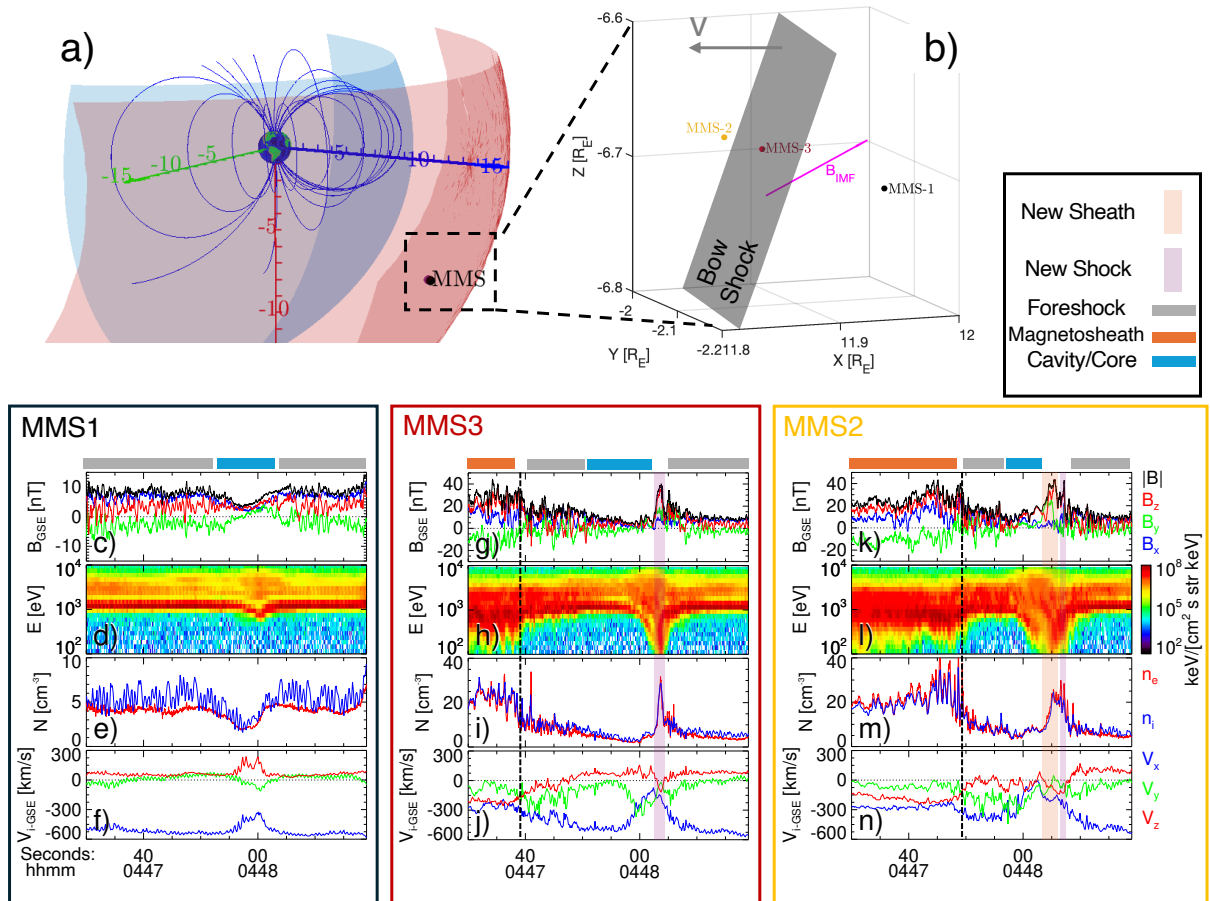

Figure 1. Overview of MMS observations of a foreshock structure on 2025-02-10. Panel (a) shows the orbit trajectory of MMS spacecraft around Earth and with respect to a model conic section bow shock (red) and magnetopause (blue). Panel (b) shows a focused view of the MMS configuration (probes 1, 2, and 3). The upstream magnetic field ($B_{IMF}$) is shown with a magenta vector, and a model bow shock plane is shown with gray, with the Sun on to the right. Panels (c - f) show MMS1 measurement timeseries of magnetic field components and strength, ion energy spectra, ion and electron densities, and ion velocities, respectively. Panels (g – j) and (k – n) show similar observables measured by MMS3 and MMS2, respectively. Observation periods corresponding to the magnetosheath (MSH), foreshock (FSH), foreshock caviton (Cav.), and core (Core) are marked above the magnetic field panels. The instances of the main bow shock, the new transient structure shock and sheath are marked by shaded boxes. Vector quantities are in the Geocentric Solar Ecliptic (GSE) coordinates

centered at Earth with the $+x$ axis pointing towards the sun and the $+y$ axis in the ecliptic plane and pointing opposite to Earth's planetary motion.

Figures 1c-f show MMS1 measurements of plasma and fields in the foreshock after MMS1 partial bow shock crossing at 04:46:59 UT. There is a correlated decrease in the magnetic field and plasma density in the middle of the shown interval and indicated by the blue bar above panel c. The magnetic field decreases by $\Delta B/B \sim$ 0.55-0.6, depending on whether the comparison is made with the IMF far upstream or with the magnetic field in the immediate vicinity. The caviton core is accompanied by an increased flux of suprathermal ions above the solar wind beam energy at 1850 eV as shown in Figure 1d. Oscillations in the total ion density in Figure 1e are in part due to foreshock backstreaming ions, small amplitude compressive foreshock wave fluctuations, and the spacecraft spin effects. The ion velocities in Figure 1f show slowdown and deflection of the solar wind core proton beam. No notable compressional rims are present around the caviton. The foreshock ULF wave activity decreases inside the core, while the IMF vectors at the upstream and downstream sides of the caviton remain the same. However, the field rotation inside the core, through a decrease mostly in $B_x$, changes the IMF cone angle within the core and increases the shock shock inclination towards quasi-perpendicular. Similar signatures are also reported in simulations of foreshock cavitons (Blanco-Cano et al., 2009). As such, the plasma and magnetic field depletion observed by MMS1 in the foreshock is likely associated with the initial stage of a ULF wave caviton.

In panels g – j of Figure 1 MMS3 data are shown for the same period. MMS3 is initially inside the magnetosheath and crosses the bow shock at 04:47:39 UT. Due to its proximity to the bow shock, MMS3 magnetic field data show a significantly higher wave activity in the foreshock. The foreshock is followed by a gradual decrease in the magnetic field strength and plasma density corresponding to the core of the foreshock caviton. The purple shaded area highlights a new compressive shock layer ("new shock") that has formed on the upstream side of the caviton. The magnetic field fluctuations upstream of the new shock increases to levels observed upstream of the main bow shock indicating that the new shock is creating its own foreshock perturbations upstream of the bow shock. The ion energy spectrogram in Figure 1h also indicates an increased flux of suprathermal ions throughout the foreshock and inside the core. In addition, heated and scattered ions within the new shock layer resemble the magnetosheath plasma observed downstream of the main bow shock with significant plasma flow deflection seen in ion velocities in Figure 1j, consistent with characteristics of a foreshock transient core region.

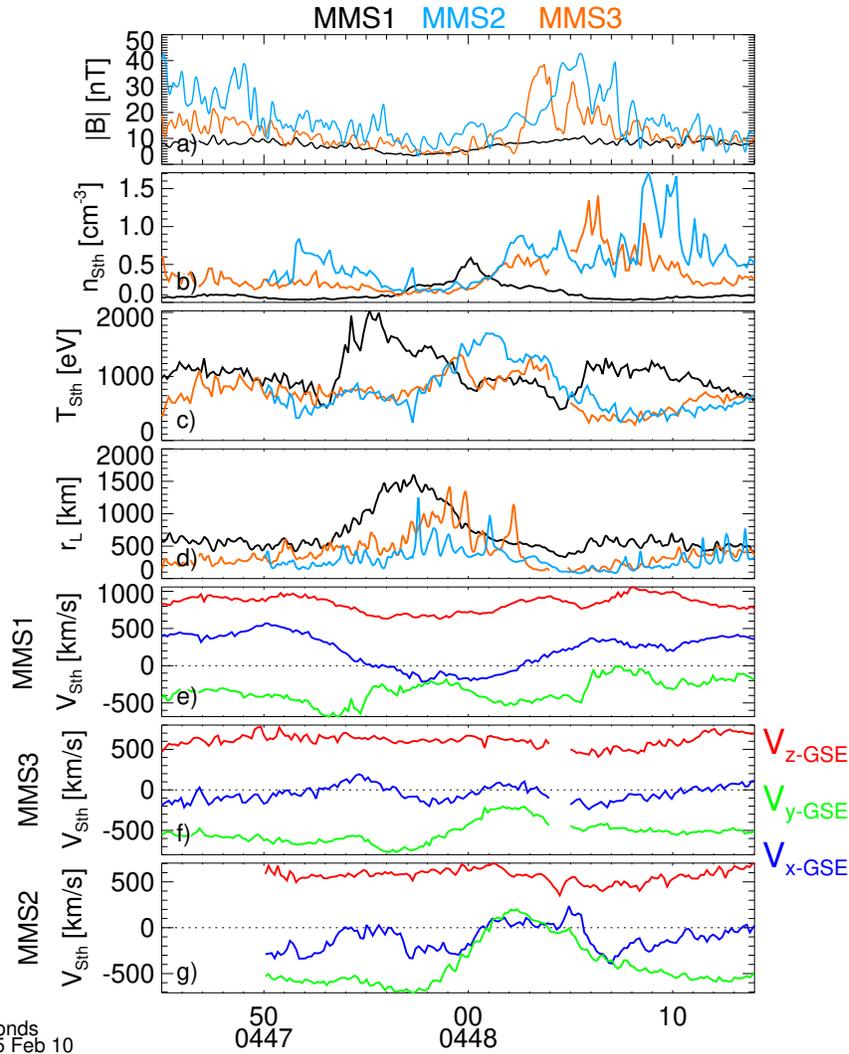

Figure 2. The suprathermal ion dynamics. (a) magnetic field profiles from MMS1 (black), MMS2 (blue), and MMS3 (orange), (b) densities, (c) average temperatures, and (d) average thermal gyroradii of suprathermal ions. Panels e – g show bulk plasma flow velocity in GSE coordinates of the suprathermal ions from MMS1, MMS3, and MMS2 respectively.

Further downstream, the MMS2 spacecraft is also initially inside the magnetosheath and crosses the bow shock at 04:47:49 UT. From the bow shock crossings by the three MMS spacecraft, we obtain an average receding bow shock speed of $V_{Bsh}$ = 19 km/s. A similar sequence of events as described for MMS3 is observed in MMS2 data. Ion energy spectra in Figure 1l indicate higher fluxes of suprathermal ions than MMS3 in the foreshock surrounding the caviton and in the core region. MMS2 measurements also indicate that the new sheath region of the foreshock transient (highlighted by red shaded area) has expanded in duration compared to MMS3 measurements. This could be due to a slower transition of the foreshock transient across the spacecraft. However, MMS2 also measures a lower plasma compression across the new sheath and shock regions of the foreshock transient, suggesting that the sheath formation is likely due to the expansion of the plasma pileup within the new shock as observed earlier by MMS3. Using the times of the highest magnetic field at MMS3 and MMS2 across the new shock we obtain a speed of $V_{Sh-new}$ ~ 128 km/s in the spacecraft frame for the new shock cycle, indicating that near the position of MMS2, the new shock has Mach number of $M_f$ ~ 4.7. The new shock is dominated by increases along $B_z$ and $B_y$ and the normal vector is mostly along the $x_{GSE}$ component.

The ion energy spectrograms in Figure 1 indicate that non-solar wind ions play a critical role in formation of a new shock cycle. For this period, we calculate the plasma moments for suprathermal ions by masking the solar wind beam ions in the 3D ion distributions and limiting the field-of-view to those occupied by suprathermal ions (Madanian et al., 2024). Figure 2a shows smoothed (0.15 s window) magnetic field profile across the foreshock transient from MMS1, MMS2, and MMS3. The MMS1 spacecraft measures the highest density of suprathermal ions near the upstream edge of the core as shown in panel (b). However, the suprathermal ions have the highest temperature earlier in the interval. The drop in the magnetic field strength increases the thermal gyroradius of these ions by as much as a factor of ~ 3. As expected, backstreaming ions that reach the MMS1 spacecraft have a net sunward flow except within the core, as shown in Figure 2e. MMS3 and MMS2 also measure higher fluxes of suprathermal ions across the new shock as compared to those within the transient core. The highest suprathermal ion densities are present upstream of the new shock structure, likely due to scattering of the solar wind ions by perturbations in the new foreshock. As the new shock develops, the net flow of suprathermal ions becomes mainly in the plane perpendicular to the solar wind flow direction (Figures 2f and g,) indicative of the back and forth motion of these ions between the main bow shock and the downstream edge of the new shock. In calculating the partial moments, we note that as the caviton core and upstream shock develop, perturbations scatter and weakened the solar wind beam while locally accelerating the scattered ions at different initial energies. As such, it becomes increasingly uncertain to separate and isolate the solar wind ions from suprathermal backstreaming ions particularly at, and upstream of the new shock where magnetic perturbations are high (see the gap in MMS3 data in Figure 2). The suprathermal ion density peaks upstream of the new shock in MMS3 and MMS2 data are caused by local scattering and reflection of solar wind ions.

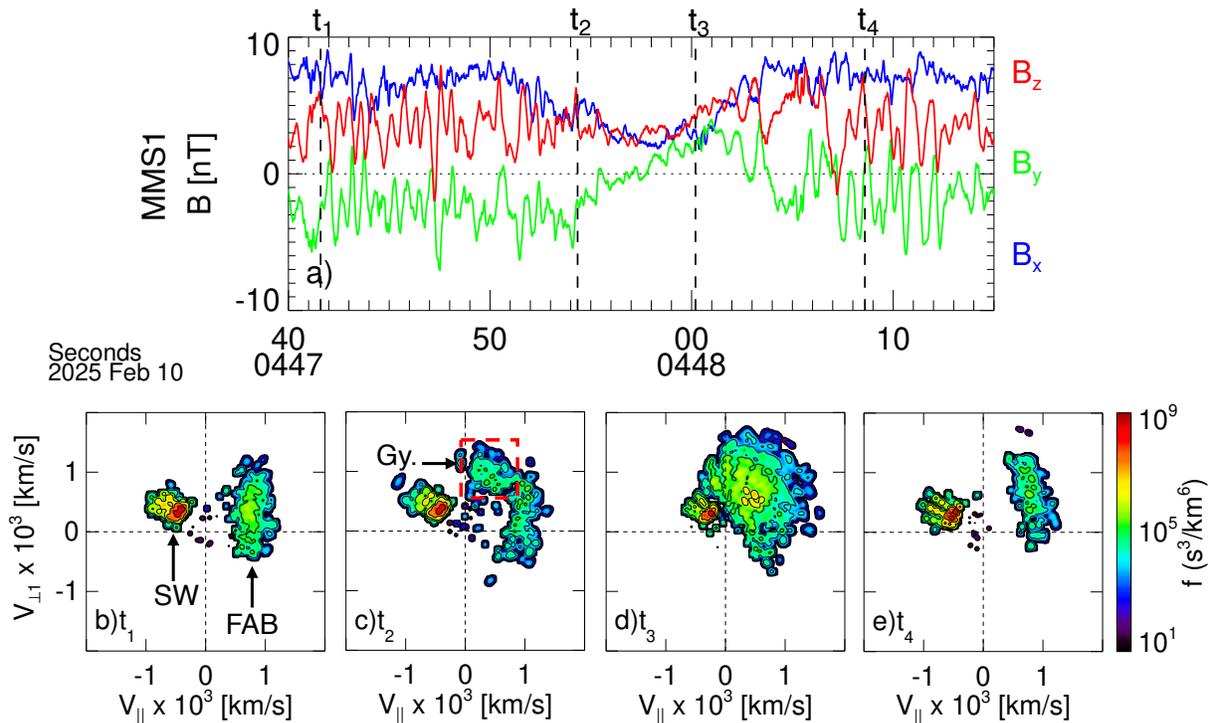

Figure 3. Ion distribution across the foreshock caviton measured by MMS1. Panels a and b show timeseries of the magnetic field vectors and suprathermal ion densities. Panels (b) through (e) show ion distributions at four timestamps in the $V_{\parallel} - V_{\perp 1}$ plane. The solar wind ions (SW), FAB, are labeled. The red box in panel (c) indicates the gyrating ion population.

The FPI instrument on MMS measures 3D distributions of positively charged particles with energies between 2.1 eV and 17.8 keV every 150 ms (Pollock et al., 2016). We analyze individual ion distributions across the caviton and the new shock to investigate the ion energetics and evolution of different ion populations. We start with MMS1 distributions in Figure 3 to determine changes in the ion distributions due to the changing shock geometry. The magnetic field vectors are shown in panel (a) as a reference with the timestamp of each distribution marked with a vertical line. Distributions are rotated to the $V_{\parallel} - V_{\perp 1}$ plane, where $V_{\parallel}$ is along the local magnetic field and $V_{\perp 1}$ is along the component of the local bulk plasma flow velocity perpendicular to the magnetic field. $V_{\perp 2}$ (not shown) completes the right-hand triple. The ion distribution in panel (b) is measured downstream of the caviton core and shows FAB ions in addition to the solar wind (SW) beam. In panel (c), another ion population, highlighted with a red box, at higher pitch angles with large $V_{\perp 1}$ emerges. This population does not follow the well-known crescent or kidney-shaped pattern seen with intermediate distributions and emerges with the onset of quasi-perpendicular shock geometry. These ions have limited gyrophase and are likely gyrating ions reflected from the quasi-perpendicular bow shock and accelerated locally before entering the caviton. However, coexistence of the two populations in Figure 3c is indicative of a not fully magnetized ion motion, perhaps due to experiencing a larger gyroradii in the core by newly accelerated ions. The abundance of gyrating ions increases within the core, as seen in distribution $t_3$ in panel (d) when the highest flux of suprathermal ions is measured. A small fraction of suprathermal ions with $-V_{\parallel}$ velocity are associated with suprathermal ions propagating anti-sunward and trapped and mirror reflected by the higher magnetic fields at the upstream edge of the caviton. High energy gyrating ions that can reach the upstream end of the caviton form the FAB ion population seen in Figure 3e.

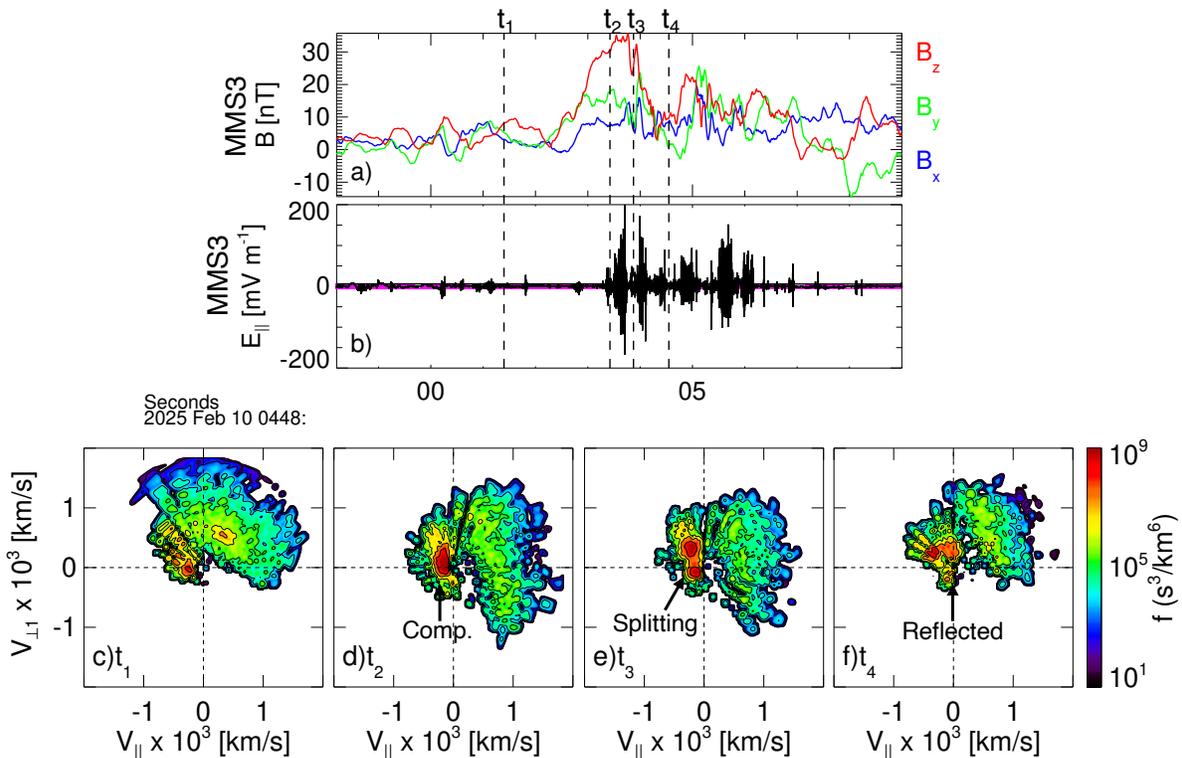

Figure 4. MMS3 measurements across the new shock at the upstream edge of the foreshock caviton. Panel (a) shows the timeseries of the magnetic field vectors. The electric field component along the background magnetic field is shown in panel b. Ion distributions at four timestamps are shown in panels (c) through (f).

While MMS1 data indicates no significant increase in the magnetic field strength or plasma density around the caviton, during travel from MMS1 to MMS3, a compressive shock layer ($B_{Max.}/B_{SW}$= 5.1, $n_{Max.}/n_{SW}$=7.4) forms at the upstream edge of the core. Figure 4 shows ion distributions measured by MMS3 across the new shock. The shock layer is accompanied by high amplitude electric fields as shown in Figure 4b. The distribution at time $t_1$ shows ions near the core-side of the new shock ramp where we see slowed down and tenuous solar wind beam ions and gyrating ions. These ions are non gyrotropic indicating a diffuse ion distribution has not formed yet at this stage. The caviton core at MMS3 also shows more depletion in the magnetic field strength and plasma density compared to MMS1 (see Figure 1e and i). Figure 4d shows ion populations at time $t_2$ at the highest magnetic field and plasma compression measured by MMS3. The flux of solar wind beam ions near the origin (i.e., slowed down beam) has intensified which explains the density compression seen in Figure 1i. This enhancement continues until the beam splits, as shown in Figure 4e, producing a new population of reflected ions. The $t_4$ distribution in panel (f) is measured upstream of the new shock and includes two populations of gyrating ions: one originated from the main bow shock at higher velocities (energies), and a newly reflected ion population from the new shock at lower energies.

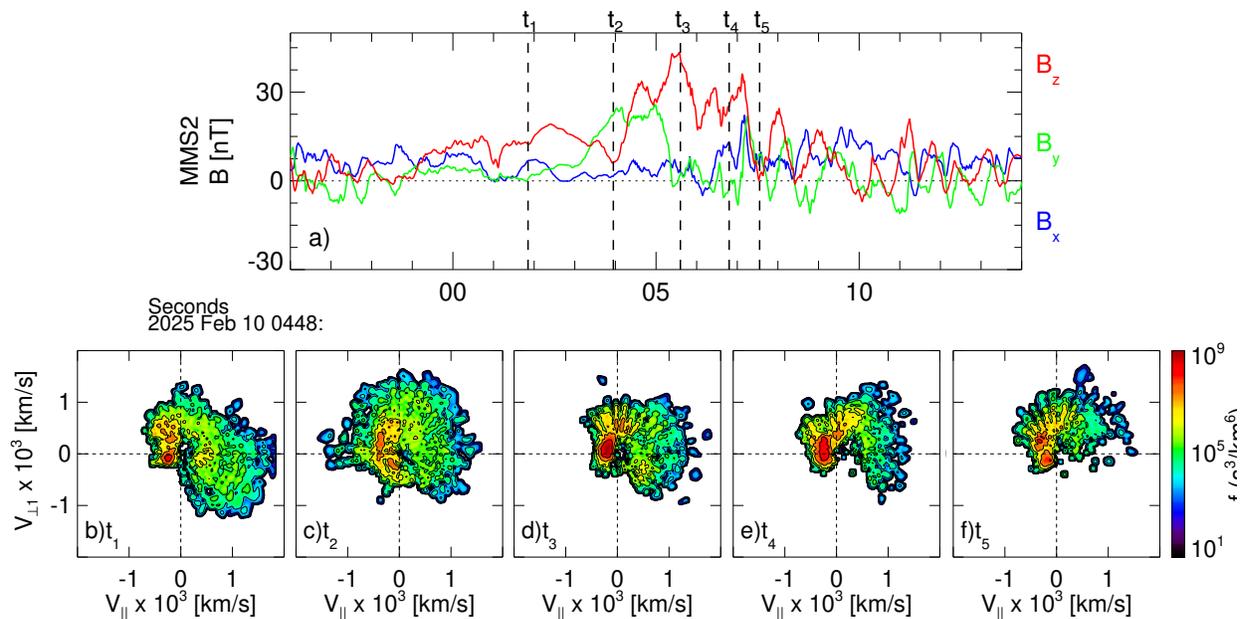

Figure 5. MMS2 measurements across the new sheath and shock. Panel (a) show the magnetic field components and strengths. Panels (b) through (f) show ion distributions at times $t_1$ through $t_5$.

The suprathermal ion densities in Figure 2b indicate that MMS2 measures higher suprathermal ion densities across the foreshock transient than MMS1 and MMS3. The ion distribution in Figure 5b shows a population of high energy gyrating ions within the core. Aside from a weakened solar wind beam, this distribution is very similar to ion distributions seen downstream of the main bow shock. Near the ramp, we begin to see isotropic diffuse ion distributions in Figure 5c. The MMS2 $t_2$ distribution in Figure 5c is measured at the same time (within a 150 ms window) as the MMS3 t₃ distribution in Figure 4e. That is when MMS3 is on the upstream ramp of the new shock while MMS2 is downstream and on the shock ramp on the depleted core side. In addition to changes to the core solar wind beam discussed earlier, comparison of these two distributions indicate that the rather diffuse ion distribution downstream of the shock is supported by anti-parallel flux of suprathermal ions and locally scattered and reflected ions which are absent in upstream of the shock. MMS2 t₃ and t₄ distributions within the extended sheath indicate lower temperatures of suprathermal ions compared to t₂ and t₃ distributions measured by MMS3 across the new shock, consistent with earlier results in Figure 2. At this time, the main bow shock is already replaced by the new shock with a quasi-perpendicular geometry and most suprathermal ions are generated by local reflection and acceleration of the solar wind ions by the motional electric field, i.e., shock drift acceleration.

## 3 Discussion

Identifying the scales and ion dynamics in in-situ measurements of supercritical quasi-parallel shocks are often complicated by the plethora of wave-particle interactions and plasma structures in this environment. By using multi-point MMS observations in the string-of-pearl configuration we analyze the reformation process of a quasi-parallel shock from a foreshock transient. We identify the initial caviton formation in the ULF wave foreshock by MMS1, the nonlinear growth of a new shock by MMS3, and the sheath expansion stage by MMS2. The foreshock caviton which leads to reformation of the shock layer is accompanied by a flux of backstreaming suprathermal ions. Due to the lack of a shear angle across the core and only an intermittent change in the cone angle, the foreshock depletion observed in MMS1 is the initial stage of a foreshock caviton, and with the upstream shock layer that develops at later stages, it essentially becomes a SHFA.

Figure 2d indicates that the average thermal gyroradius of suprathermal ions within the caviton increases. MMS1 ion distributions reveal that a flux of gyrating ions emerge within the core as the shock geometry changes to quasi-perpendicular. High energy gyrating ions that reach the upstream side of the caviton contribute to the FAB ions. They also create a cross field current leading to instabilities, such as modified two stream instability, that manifest as broadband field aligned electrostatic waves, which are present in Figure 2b, and nonlinear growth of the magnetic field. Similar ion gyrokinetic effects at high beta quasi-perpendicular shocks also lead to high magnetic amplifications (Madanian et al., 2021). Decreases in the average ion gyroradii within the caviton core from ~ 1500 to < 300 km in Figure 2d, and the expansion of the sheath region with enhanced magnetic fields ($V_{Sh-new} * \Delta t \sim 770$ km, where $\Delta t$ is the duration of the new sheath and shock regions in Figure 1k) suggest that as the new shock and its sheath layer expand, they can prevent foreshock ions from penetrating and appearing upstream suppressing the cross-field currents.

Distributions in Figure 4 indicate that electrostatic waves interact with the solar wind beam causing compactification and reflection of the beam. This process occurs within a spatial scale of 2 $l_i$ based on the new shock speed and the duration of the shock width. MMS2 data in Figure 5 further reveals that as the structure evolves in the foreshock, the pressure gradient associated with the compressed solar wind beam relaxes as the plasma expands and a new sheath develops. The new shock forms in the foreshock upstream of MMS3 and downstream of MMS1, or within a distance of 4.5 and 11.2 $l_i$ from the main bow shock. The MMS spacecraft travel at ~2 km/s along the bow shock normal during this period. The abundance of backstreaming ions decreases with distance to the bow shock. However, that may not be a critical factor for the new shock reformation, as suprathermal ion densities at MMS1 reach as high as those observed by MMS3. The fraction of suprathermal ions that reach the upstream edge of the caviton contribute to cross-field currents at that region and to the growth of instabilities. Suprathermal densities in the post new shock region do vary between MMS1 and MMS3 or MMS2 as shown in Figure 2b. The gyrating suprathermal ions in the caviton core result in cross-field currents that transfer energy through electric fields to the solar wind plasma and form the new shock. This shock formation at a foreshock transient in some ways is similar to foreshock bubble shocks (Madanian et al., 2023). A foreshock bubble's shock expansion speed can be modeled using the properties of foreshock ions (Liu et al., 2023). If we assume the field depletion across the ULF caviton act as an embedded discontinuity in the upstream plasma across which cross-field currents by the foreshock ions are generated, we can use the foreshock-to-solar wind ion density ratio of ~ 20% (Figures 1e and 2b,) and the foreshock ion gyrovelocity of ~ 500 km/s (Figure 4c) in Eq. 3 from (Liu et al., 2023) to estimate a shock speed of ~ 353 km/s (~ 116 km/s) in the solar wind (spacecraft) rest frame, consistent with the estimated value from the timing analysis in Section 2. This agreement lends further support to the shock reformation process described above.

The shock reformation process in this study arises from ion kinetic scale processes. The shock front can also reform on smaller scales due to electron scale current sheets and whistler waves (Krasnoselskikh et al., 2002; Turner et al., 2021). Shock reformation can occur simultaneously on multiple scales ranging from the electron-kinetic to ion-kinetic scales leading to the presence of a global-scale shock layer. The new shock layer that forms in the foreshock can be viewed as a hybrid shock layer. That is, not all ions interact with the electromagnetic fields at the shock front in the same way or interact at all. Backstreaming ions reflected at the main bow shock can pass through the shock and appear upstream of the new shock, while fluxes of locally reflected ions at the new shock appear both upstream in the solar wind, and downstream within the core. In the upstream, these ions are quickly accelerated by the motional electric field, however, within the core, they form a low energy pitch-angle scattered ion population. In addition, the new shock, and most other perturbations in the foreshock region, tends to have a quasi-perpendicular orientation. However, as they merge with the bow shock the quasi-parallel shock orientation will reestablish. Shock reformation is enabled by transient shift in the shock geometry from quasi-parallel to quasi-perpendicular regime.

Gyrating ions can originate from different parts of the entire curved bow shock surface. Shock simulations in 1D or 2D space without a full 3D surface morphology may not capture the full contribution of gyrating ions and hence underestimate the reformation rate. We also note that based on the shock angle and the point of observations in Figure 1a, MMS spacecraft are near the foreshock boundary where lateral expansion of foreshock structures developed in the middle of the foreshock region can be observed intermittently (Omidi, Sibeck, et al., 2013).

In our analysis of suprathermal ion distributions and partial moments we assume a purely proton plasma. However, alpha particles have been observed in the composition of backstreaming ions with abundances comparable to those of the upstream solar wind plasma (Burgess, 1989; Fuselier et al., 1995). The alpha particle to proton density ratio in the upstream solar wind during the event we analyzed is ~ 6%, and some ion fluxes in our observations could be alpha particles. The potential presence of a small fraction of alpha particles could slightly change the reported suprathermal ion moments in Figure 2, however, it will not affect our interpretations of observations.

## 4 Conclusions

In this study we use multi-point measurements by the MMS spacecraft to investigate a reformation cycle of a supercritical quasi-parallel shock. The cycle begins at the upstream edge of a foreshock caviton generated by the flux of backstreaming ions and cross mode interactions of the ULF waves. The magnetic field depression changes the bow shock geometry towards a quasi-perpendicular regime while at the same time increasing the backstreaming ions gyroradius. Our analysis shows that as the foreshock caviton approaches the bow shock and evolves into an SHFA, a compressional shock layer is formed at the upstream edge of the core accompanied by compression and compactification of the incident upstream cold proton beam. The expansion of this layer contributes to the new sheath layer ahead of the main bow shock. This reformation process occurs repeatedly as the reformed bow shock regains quasi-parallel geometry and backstreaming ions cause ULF waves.

The spatiotemporal limits for the shock growth reported in this study can be used as a benchmark for numerical simulation of foreshock structures. The results also signify the role of upstream structures in reformation of quasi-parallel supercritical shocks particularly when the ULF wave steepening into SLAMS is not strong enough to cause quasi-periodic reformation, but rather cavitons form which evolve into SHFA shocks and reform the main bow shock.

**Data and Code Availability**

Link to ion distribution dataset: https://doi.org/10.48322/dq1y-nf73, magnetic field dataset: https://doi.org/10.48322/pj0n-m695, and electric field dataset: https://doi.org/10.48322/zn13-qd40. Link to the software package to retrieve, analyze, and visualize data: https://spedas.org/, and link to an example routine to regenerate Figures 1.c-f: https://github.com/.

**References**

An, X., Liu, T. Z., Bortnik, J., Osmane, A., & Angelopoulos, V. (2020). Formation of Foreshock Transients and Associated Secondary Shocks. *The Astrophysical Journal*, *901*(1), 73. https://doi.org/10.3847/1538-4357/abaf03

Blanco-Cano, X., Omidi, N., & Russell, C. T. (2009). Global hybrid simulations: Foreshock waves and cavitons under radial interplanetary magnetic field geometry. *Journal of Geophysical Research: Space Physics*, *114*(A1), 2008JA013406. https://doi.org/10.1029/2008JA013406

Burgess, D. (1989). Alpha particles in field-aligned beams upstream of the bow shock: Simulations. *Geophysical Research Letters*, *16*(2). https://doi.org/10.1029/GL016i002p00163

Eastwood, J. P., Lucek, E. A., Mazelle, C., Meziane, K., Narita, Y., Pickett, J., & Treumann, R. A. (2005). The foreshock. *Space Science Reviews*, *118*(1–4), 41–94. https://doi.org/10.1007/s11214-005-3824-3

Fuselier, S. A. (1995). Ion distributions in the Earth's foreshock upstream from the bow shock. *Advances in Space Research*, *15*(8–9), 43–52. https://doi.org/10.1016/0273-1177(94)00083-D

Fuselier, S. A., Thomsen, M. F., Ipavich, F. M., & Schmidt, W. K. H. (1995). Suprathermal He 2+ in the Earth's foreshock region. *Journal of Geophysical Research*, *100*(A9), 17107. https://doi.org/10.1029/95ja00898

Gary, S. P. (1991). Electromagnetic ion/ion instabilities and their consequences in space plasmas: A review. *Space Science Reviews*, *56*(3–4). https://doi.org/10.1007/BF00196632

Hao, Y., Gao, X., Lu, Q., Huang, C., Wang, R., & Wang, S. (2017). Reformation of rippled quasi-parallel shocks: 2-D hybrid simulations. *Journal of Geophysical Research: Space Physics*, *122*(6), 6385–6396. https://doi.org/10.1002/2017JA024234

Hoppe, M. M., Russell, C. T., Frank, L. A., Eastman, T. E., & Greenstadt, E. W. (1981). Upstream hydromagnetic waves and their association with backstreaming ion populations: ISEE 1 and 2 observations. *Journal of Geophysical Research: Space Physics*, *86*(A6), 4471–4492. https://doi.org/10.1029/JA086iA06p04471

Johlander, A., Battarbee, M., Turc, L., Ganse, U., Pfau-Kempf, Y., Grandin, M., Suni, J., Tarvus, V., Bussov, M., Zhou, H., Alho, M., Dubart, M., George, H., Papadakis, K., & Palmroth, M. (2022). Quasi-Parallel Shock Reformation Seen by Magnetospheric Multiscale and Ion-Kinetic Simulations. *Geophysical Research Letters*, *49*(2), e2021GL096335. https://doi.org/10.1029/2021GL096335

Kajdič, P., Hietala, H., & Blanco-Cano, X. (2017). Different Types of Ion Populations Upstream of the 2013 October 8 Interplanetary Shock. *The Astrophysical Journal Letters*, *849*(2), L27. https://doi.org/10.3847/2041-8213/aa94c6

Kempf, Y., Pokhotelov, D., Gutynska, O., Wilson Iii, L. B., Walsh, B. M., Alfthan, S. V., Hannuksela, O., Sibeck, D. G., & Palmroth, M. (2015). Ion distributions in the Earth's foreshock: Hybrid-Vlasov simulation and THEMIS observations. *Journal of Geophysical Research: Space Physics*, *120*(5), 3684–3701. https://doi.org/10.1002/2014JA020519

Krasnoselskikh, V., Lembège, B., Savoini, P., & Lobzin, V. V. (2002). Nonstationarity of strong collisionless quasiperpendicular shocks: Theory and full particle numerical simulations. *Physics of Plasmas*, *9*(4), 1192–1209. https://doi.org/10.1063/1.1457465

Liu, T. Z., An, X., Zhang, H., & Turner, D. (2020). Magnetospheric Multiscale Observations of Foreshock Transients at Their Very Early Stage. *The Astrophysical Journal*, *902*(1), 5. https://doi.org/10.3847/1538-4357/abb249

Liu, T. Z., Hietala, H., Angelopoulos, V., & Turner, D. L. (2016). Observations of a new foreshock region upstream of a foreshock bubble's shock. *Geophysical Research Letters*, *43*(10), 4708–4715. https://doi.org/10.1002/2016GL068984

Liu, T. Z., Vu, A., Zhang, H., An, X., & Angelopoulos, V. (2023). Modeling the Expansion Speed of Foreshock Bubbles. *Journal of Geophysical Research: Space Physics*, *128*(2), e2022JA030814. https://doi.org/10.1029/2022JA030814

Madanian, H., Desai, M. I., Schwartz, S. J., Wilson, L. B., Fuselier, S. A., Burch, J. L., Contel, O. L., Turner, D. L., Ogasawara, K., Brosius, A. L., Russell, C. T., Ergun, R. E., Ahmadi, N., Gershman, D. J., & Lindqvist, P.-A. (2021). The Dynamics of a High Mach Number Quasi-perpendicular Shock: MMS Observations. *The Astrophysical Journal*, *908*(1), 40. https://doi.org/10.3847/1538-4357/abcb88

Madanian, H., Gingell, I., Chen, L.-J., & Monyek, E. (2024). Drivers of Magnetic Field Amplification at Oblique Shocks: In Situ Observations. *The Astrophysical Journal Letters*, *965*(1), L12. https://doi.org/10.3847/2041-8213/ad3073

Madanian, H., Liu, T. Z., Phan, T. D., Trattner, K. J., Karlsson, T., & Liemohn, M. W. (2022). Asymmetric Interaction of a Solar Wind Reconnecting Current Sheet and Its Magnetic Hole With Earth's Bow Shock and Magnetopause. *Journal of Geophysical Research: Space Physics*, *127*(4). https://doi.org/10.1029/2021JA030079

Madanian, H., Omidi, N., Sibeck, D. G., Andersson, L., Ramstad, R., Xu, S., Gruesbeck, J. R., Schwartz, S. J., Frahm, R. A., Brain, D. A., Kajdic, P., Eparvier, F. G., Mitchell, D. L., & Curry, S. M. (2023). Transient Foreshock Structures Upstream of Mars: Implications of the Small Martian Bow Shock. *Geophysical Research Letters*, *50*(8), e2022GL101734. https://doi.org/10.1029/2022GL101734

Meziane, K., Wilber, M., Mazelle, C., LeQuéau, D., Kucharek, H., Lucek, E. A., Rème, H., Hamza, A. M., Sauvaud, J. A., Bosqued, J. M., Dandouras, I., Parks, G. K., McCarthy, M., Klecker, B., Korth, A., Bavassano-Cattaneo, M. B., & Lundin, R. N. (2004). Simultaneous observations of field-aligned beams and gyrating ions in the terrestrial foreshock. *Journal of Geophysical Research: Space Physics*, *109*(A5), 2003JA010374. https://doi.org/10.1029/2003JA010374

Omidi, N., Karimabadi, H., Krauss-Varban, D., & Killen, K. (1994). Generation and nonlinear evolution of oblique magnetosonic waves: Application to foreshock and comets. In J. L. Burch & J. H. Waite (Eds.), *Geophysical Monograph Series* (Vol. 84, pp. 71–84). American Geophysical Union. https://doi.org/10.1029/GM084p0071

Omidi, N., Sibeck, D., Blanco-Cano, X., Rojas-Castillo, D., Turner, D., Zhang, H., & Kajdič, P. (2013). Dynamics of the foreshock compressional boundary and its connection to foreshock cavities. *Journal of Geophysical Research: Space Physics*, *118*(2), 823–831. https://doi.org/10.1002/jgra.50146

Omidi, N., Zhang, H., Sibeck, D., & Turner, D. (2013). Spontaneous hot flow anomalies at quasi-parallel shocks: 2. Hybrid simulations. *Journal of Geophysical Research: Space Physics*, *118*(1), 173–180. https://doi.org/10.1029/2012JA018099

Parks, G. K., Lee, E., Mozer, F., Wilber, M., Lucek, E., Dandouras, I., Rème, H., Mazelle, C., Cao, J. B., Meziane, K., Goldstein, M. L., & Escoubet, P. (2006). Larmor radius size density holes discovered in the solar wind upstream of Earth's bow shock. *Physics of Plasmas*, *13*(5), 050701. https://doi.org/10.1063/1.2201056

Pollock, C., Moore, T., Jacques, A., Burch, J., Gliese, U., Saito, Y., Omoto, T., Avanov, L., Barrie, A., Coffey, V., Dorelli, J., Gershman, D., Giles, B., Rosnack, T., Salo, C., Yokota, S., Adrian, M., Aoustin, C., Auletti, C., … Zeuch, M. (2016). Fast Plasma Investigation for Magnetospheric Multiscale. *Space Science Reviews*, *199*(1–4), 331–406. https://doi.org/10.1007/s11214-016-0245-4

Savoini, P., & Lembège, B. (2015). Production of nongyrotropic and gyrotropic backstreaming ion distributions in the quasi-perpendicular ion foreshock region. *Journal of Geophysical Research: Space Physics*, *120*(9), 7154–7171. https://doi.org/10.1002/2015JA021018

Schwartz, S. J. (1998). Shock and Discontinuity Normals, Mach Numbers, and Related Parameters. In G. Paschmann & P. W. Daly (Eds.), *Analysis Methods for Multi-Spacecraft Data (ISSI Scientific Reports Series)* (Vol. 1, pp. 249–270).

Schwartz, S. J., Burgess, D., Wilkinson, W. P., Kessel, R. L., Dunlop, M., & Lühr, H. (1992). Observations of short large-amplitude magnetic structures at a quasi-parallel shock. *Journal of Geophysical Research*, *97*(A4), 4209. https://doi.org/10.1029/91JA02581

Sibeck, D. G., Phan, T. D., Lin, R., Lepping, R. P., & Szabo, A. (2002). Wind observations of foreshock cavities: A case study. *Journal of Geophysical Research: Space Physics*, *107*(A10). https://doi.org/10.1029/2001JA007539

Stasiewicz, K., & Kłos, Z. (2022). On the formation of quasi-parallel shocks, magnetic and electric field turbulence, and the ion energization mechanism. *Monthly Notices of the Royal Astronomical Society*, *513*(4), 5892–5899. https://doi.org/10.1093/mnras/stac1193

Tarvus, V., Turc, L., Battarbee, M., Suni, J., Blanco-Cano, X., Ganse, U., Pfau-Kempf, Y., Alho, M., Dubart, M., Grandin, M., Johlander, A., Papadakis, K., & Palmroth, M. (2021). Foreshock cavitons and spontaneous hot flow anomalies: A statistical study with a global hybrid-Vlasov simulation. *Annales Geophysicae*, *39*(5), 911–928. https://doi.org/10.5194/angeo-39-911-2021

Thomsen, M. F., Gosling, J. T., Fuselier, S. A., Bame, S. J., & Russell, C. T. (1986). Hot, diamagnetic cavities upstream from the Earth's bow shock. *Journal of

*Geophysical Research: Space Physics*, *91*(A3), 2961–2973. https://doi.org/10.1029/JA091iA03p02961

Trattner, K. J., Fuselier, S. A., Schwartz, S. J., Kucharek, H., Burch, J. L., Ergun, R. E., Petrinec, S. M., & Madanian, H. (2023). Ion Acceleration at the Quasi-Parallel Shock: The Source Distributions of the Diffuse Ions. *Journal of Geophysical Research: Space Physics*, *128*(2), e2022JA030631. https://doi.org/10.1029/2022JA030631

Turner, D. L., Wilson, L. B., Goodrich, K. A., Madanian, H., Schwartz, S. J., Liu, T. Z., Johlander, A., Caprioli, D., Cohen, I. J., Gershman, D., Hietala, H., Westlake, J. H., Lavraud, B., Le Contel, O., & Burch, J. L. (2021). Direct Multipoint Observations Capturing the Reformation of a Supercritical Fast Magnetosonic Shock. *The Astrophysical Journal Letters*, *911*(2), L31. https://doi.org/10.3847/2041-8213/abec78

Vu, A., Liu, T. Z., Zhang, H., & Delamere, P. (2023). Parameter Dependencies of Early-Stage Tangential Discontinuity-Driven Foreshock Bubbles in Local Hybrid Simulations. *Journal of Geophysical Research: Space Physics*, *128*(2), e2022JA030815. https://doi.org/10.1029/2022JA030815

Wang, S., Bessho, N., Graham, D. B., Le Contel, O., Wilder, F. D., Khotyaintsev, Y. V., Genestreti, K. J., Lavraud, B., Choi, S., & Burch, J. L. (2022). Whistler Waves Associated With Electron Beams in Magnetopause Reconnection Diffusion Regions. *Journal of Geophysical Research: Space Physics*, *127*(9), e2022JA030882. https://doi.org/10.1029/2022JA030882

Wilson, L. B., Koval, A., Sibeck, D. G., Szabo, A., Cattell, C. A., Kasper, J. C., Maruca, B. A., Pulupa, M., Salem, C. S., & Wilber, M. (2013). Shocklets, SLAMS, and field-aligned ion beams in the terrestrial foreshock. *Journal of Geophysical Research: Space Physics*, *118*(3), 957–966. https://doi.org/10.1029/2012JA018186

Winske, D., & Leroy, M. M. (1984). Diffuse ions produced by electromagnetic ion beam instabilities. *Journal of Geophysical Research: Space Physics*, *89*(A5), 2673–2688. https://doi.org/10.1029/JA089iA05p02673

Zhang, H., Sibeck, D. G., Zong, Q. -G., Omidi, N., Turner, D., & Clausen, L. B. N. (2013). Spontaneous hot flow anomalies at quasi-parallel shocks: 1. Observations. *Journal of Geophysical Research: Space Physics*, *118*(6), 3357–3363. https://doi.org/10.1002/jgra.50376

Zhu, X., Wang, M., Shi, Q., Zhang, H., Tian, A., Yao, S., Guo, R., Liu, J., Bai, S., Degeling, A. W., Zhang, S., Niu, Z., Zhao, J., Xiao, Y., & Shang, W. (2021). Motion of Classic and Spontaneous Hot Flow Anomalies Observed by Cluster. *Journal of Geophysical Research: Space Physics*, *126*(11), e2021JA029418. https://doi.org/10.1029/2021JA029418